# From Multi-sig to DLCs: Modern Oracle Designs on Bitcoin

Giulio Caldarelli
*University of Turin*
giulio.caldarelli@unito.it
**ORCID-ID: 0000-0002-8922-7871**

## Abstract

Unlike Ethereum, which was conceived as a general-purpose smart-contract platform, Bitcoin was designed primarily as a transaction ledger for its native currency, which limits programmability for conditional applications. This constraint is particularly evident when considering oracles, mechanisms that enable Bitcoin contracts to depend on exogenous events. This paper investigates whether new oracle designs have emerged for Bitcoin Layer 1 since the 2015 transition to the Ethereum smart contracts era and whether subsequent Bitcoin improvement proposals have expanded oracles' implementability. Using Scopus and Web of Science searches, complemented by Google Scholar to capture protocol proposals, we observe that the indexed academic coverage remains limited, and many contributions circulate outside journal venues. Within the retrieved corpus, the main post-2015 shift is from multisig-style, which envisioned oracles as co-signers, toward attestation-based designs, mainly represented by Discreet Log Contracts (DLCs), which show stronger Bitcoin community compliance, tool support, and evidence of practical implementations in real-world scenarios such as betting and prediction-market mechanisms.

## 1. Introduction

*"Bitcoin itself is a bounded oracle implementation that can produce compact proofs"* (Peter Todd [1]). Todd's remark highlights that Bitcoin can serve as an exceptionally reliable oracle for facts internal to its own consensus, such as the existence, ordering, and validity of transactions. The challenge arises when applications require extrinsic information about events and data originating outside the blockchain, while still preserving Bitcoin's security and minimizing additional trust assumptions. The problem of designing robust oracle mechanisms for Bitcoin is longstanding and can be traced back to the ecosystem's earliest efforts to support conditional transactions. One of the earliest documented proposals for a Bitcoin transaction conditional on an exogenous event appears on the Bitcoin Wiki [2]. In 2011, the developer Mike Hearn suggested addressing Bitcoin inheritance by making a transaction contingent on extrinsic information, the wallet owner's death in that case, provided by an "oracle". The basic idea is to construct an m-of-n multi-signature output in which one of the n keys is held by an arbiter (the oracle), who co-signs and thereby unlocks the transaction once a specified event outcome has been verified. Despite Nakamoto's skepticism [3] about introducing a third party due to the evident implications for the trustworthiness of data reporters, which inevitably reintroduce a single point of failure into the blockchain, these conditional transactions aroused the interest of many developers, who contributed to extending the basic concept. Oraclize, for example, leveraged trusted execution environment (TEE) technology to go beyond simple m-of-n oracle schemes by providing cryptographic attestations that allowed contracts to verify that oracle data was fetched from the declared source and executed within a trusted enclave [4]. The Orisi project, by contrast, explicitly addressed the risks associated with single-oracle trust assumptions by adopting a larger federation of oracle signers. Rather than relying on small multisignature configurations, Orisi proposed m-of-n schemes with significantly higher thresholds (e.g., 8-of-11), distributing the responsibilities of the oracle across multiple independent entities to improve fault tolerance and availability [5]. These basic systems worked well for simple applications such as bets, but implementing prediction markets, for example, proved more complex and required features that Bitcoin couldn't provide. To overcome these limitations, Truthcoin proposed shifting oracle computations to auxiliary chains while relying on the Bitcoin blockchain for final settlement [6]. Building on early sidechain concepts, the proposal aimed to preserve Bitcoin's security while enabling more complex off-chain oracle mechanisms. While the approach attracted conceptual interest within the Bitcoin community, the practical development of sidechain infrastructures proved significantly more complex and slower than initially anticipated. When more complex applications, such as decentralized exchanges, were finally proposed on Bitcoin, it became evident that continuously updated extrinsic data, most notably price feeds, could not be efficiently supported through multi-signature schemes due to cost, scalability, and network overhead. To address this limitation, the Counterparty project implemented oracle and exchange functionalities through a metaprotocol encoded in Bitcoin transactions using the OP_RETURN field [7]. However, the growing use of OP_RETURN for embedding application data sparked concerns among parts of the Bitcoin community about blockchain bloat and network sustainability, leading to pressure to limit OP_RETURN size and restrict non-transactional data inclusion [8], [9]. This debate, often labeled as "war," among other factors, pushed an early Bitcoin developer, Vitalik Buterin, to build his own chain with explicit support for smart contracts and oracles [10], [11], [12]. The advent of Ethereum in 2015 triggered a major shift in decentralized application development away from Bitcoin toward Ethereum. Oracle developers likewise gravitated toward the new platform, attracted by its expressive smart contract model, greater availability of development resources, increased funding opportunities, and a more accessible programming environment. Although the formal development of decentralized applications largely continued on Ethereum, the Bitcoin community pursued a steady evolution of the protocol after 2015 through a series of major upgrades. Most notably, Segregated Witness (SegWit) improved transaction malleability and efficiency, while Taproot introduced Schnorr signatures and enhanced script privacy, enabling more articulated and flexible spending conditions [13], [14]. Combined with later advances in off-chain solutions and contract-like techniques, these developments reopened questions about Bitcoin's capacity to support more complex applications directly on its main blockchain. Almost ten years after the shift toward Ethereum, we reassess whether Bitcoin Layer 1 oracle development, intended as the provision of external data to applications built directly on Bitcoin, has persisted, and how it has evolved.

By adopting a scoping review, this paper addresses this gap by answering the following research questions:

**RQ1.** What new design avenues for Bitcoin oracles have emerged since the launch of Ethereum in 2015?

**RQ2.** How is Bitcoin-oracle development currently pursued?

**RQ3.** Which real-world applications are the most documented/proposed on Bitcoin Layer 1, given the available oracle mechanisms?

The remainder of the paper is organized as follows. Section 2 maps the academic literature on mechanisms for bringing extrinsic data to Bitcoin. Section 3 describes the review methodology. Section 4 reports the findings. Section 5 discusses the findings in relation to the research questions. Section 6 concludes the paper and outlines directions for future research.

## 2. Literature Review

While the oracle literature remains relatively niche, several active research streams continue to contribute to it on a regular basis. These streams, however, mainly focus on Ethereum and EVM-compatible chains, while Bitcoin, given fewer direct real-world integrations, remains comparatively underexplored under the specific lens of oracles. The first review focusing on Bitcoin oracle mechanisms dates to 2023 and covers the period from Bitcoin's launch up to the advent of Ethereum [15]. The scope of that article was to trace the early development of Bitcoin oracles chronologically, highlighting technical constraints, community responses, and integration attempts. A recent study on cryptographic primitives for payment channels lists some that are used in modern oracle schemes but does not delve into their design and architecture [16]. While academic work specifically on Bitcoin L1 oracles remains limited beyond this contribution, other streams provide relevant insights on how Bitcoin interacts with extrinsic data. One active area concerns bridges, which rely on oracle-like assumptions to attest external chain state and enable cross-chain interoperability. Tang et al. [17], provide a taxonomy of Bitcoin cross-chain bridge protocols, classifying designs into naive token swapping, pegged-asset bridges, and arbitrary-message bridges, and comparing them along dimensions such as trust assumptions, latency, capital efficiency, and computational overhead.

Rollups and Bitcoin Layer-2 designs are also sometimes discussed within broader oracle taxonomies, because many architectures embed components that play an oracle-like role (e.g., attesting off-chain state before it is linked back to Bitcoin). Recent surveys of Bitcoin L2 and rollup designs also indicate that several constructions incorporate explicit oracle components within their architecture. For example, some BitVM-based designs rely on trusted attestation for specific validation steps, while other platforms integrate oracle networks and oracle-driven bridge components for state verification before committing to Bitcoin [18].

In the software architecture literature, Xu et al. [19] also formalize the concept of Reverse Oracle patterns as standard ways for external systems to consume blockchain state, where Bitcoin serves as the data provider for on-chain transactions.

Although these mechanisms relate to oracle taxonomies because they either bring external information into Bitcoin-adjacent systems or export Bitcoin state outward, this study focuses on a narrower meaning of Bitcoin L1 oracles concerning mechanisms that bring real-world data to Bitcoin L1, such as asset prices, event outcomes, or weather conditions, to enable conditional transaction execution [20]. Projects such as Reality Keys, Oraclize, and Orisi have historically represented this branch, which typically requires a reliable data source, a secure communication channel, and a robust incentive mechanism to deter collusion and manipulation.

Despite recent work on other oracle-like mechanisms, such as the recent review by Tang et al. [17] on Bitcoin bridges, there is little synthesis focusing on Bitcoin L1 oracles as extrinsic data providers, especially post-SegWit/Taproot and in light of DLC-style constructions. This study seeks to fill this gap by reviewing related literature, and the next paragraph outlines the adopted methodology.

### 3. Methodology and data collection

A previous overview of Bitcoin oracles enumerates proposals and projects from 2011 to early 2015 [15]. During that period, academic publications on blockchain and, in particular, Bitcoin oracles were scarce. Since blockchain literature is now mature, we updated the state of the art by conducting a structured search in major bibliographic databases.

We queried Scopus and Web of Science using the keywords Bitcoin AND Oracle with the field option "All fields" for both terms, to maximize inclusion. The searches were executed on 04 December 2025. Scopus returned 181 records, and Web of Science returned 96. After merging both samples, we obtained a dataset of 186 records, from which we removed 92 duplicates. We screened the titles and abstracts to identify works that:

1) propose oracle mechanisms operating under Bitcoin's design constraints, or

2) extend existing Bitcoin oracle designs with technical improvements or real-world implementations.

Works were excluded when "oracle" referred to unrelated meanings (e.g., random-oracle model, software testing oracles, database systems) or when Bitcoin was only a contextual mention. As stated in the literature background, we exclude from this review oracle-like mechanisms such as bridges and rollups. Although these mechanisms can transfer extrinsic information on-chain, they are designed around a different rationale. We also exclude Layer-2 oracles, as, despite serving the same broad function as those subjects of this study, they are not constrained by Layer-1 mechanics.

After initial screening, 166 records were excluded at abstract screening as unrelated. The remaining 19 were downloaded and assessed in full text, of which 1 was retained as directly relevant. This retained work explicitly referenced additional relevant contributions that were not captured by Scopus/Web of Science because they were disseminated outside journal venues, such as preprints or practitioner protocol proposals. This observation motivated a complementary search strategy aimed at capturing non-indexed but technically substantive contributions.

Accordingly, we complemented database searches with Google Scholar using the same keywords. We screened results up to rank 500, at which point results became dominated by duplicates and clearly unrelated materials. From this process, we shortlisted 42 documents for full-text assessment and retained 6 that directly investigate or propose Bitcoin oracle solutions. After a full read of the initial sample and systemizing the knowledge, we noted that Discreet Log Contracts emerged as a core concept for modern oracles, and that the construct "cryptographic oracle" was often used as a generic term for those serving UTXO chains, to which Bitcoin belongs. For that reason, we rerun the research on 12 Jan 2026 using queries based on the above-mentioned keywords first on Scopus/WoS and then on Google Scholar. Leveraging the same exclusion criteria, we extracted another 4 articles. Using backward snowballing, we further screened the reference lists of included items to minimize omission. The entire search strategy is outlined in Figure 1, which is based on the PRISMA statement model [21]. Overall, the final corpus comprises eleven documents, summarized in Table 1. The small number of eligible studies supports the view that Bitcoin-native oracle mechanisms remain underexplored in the academic literature, with several key contributions appearing outside traditional journal venues.

**Figure 1.** Review Methodology

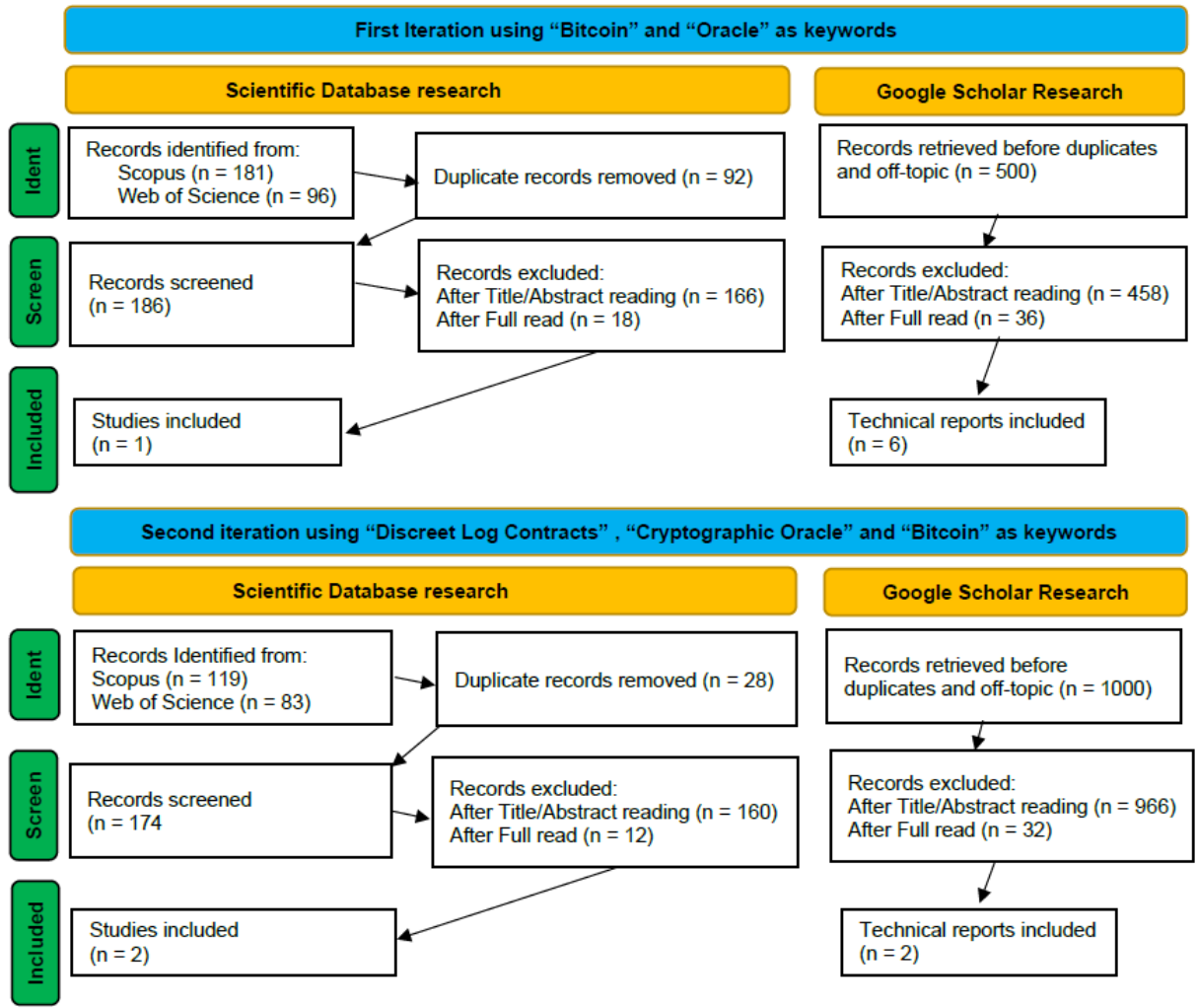

**Table 1.** Included contributions on modern Bitcoin oracles

| Date | Author | Title | Source type | Contribution |
|------|--------|-------|-------------|--------------|
| 2016 | Wright, C. [22] | Turing Complete Bitcoin Script | Whitepaper | Proposes an off-chain automation oracle/bot that repeatedly interacts with Bitcoin Script to emulate iterative behavior (loops). |
| 2017 | Dryja, T. [23] | Discreet Log Contracts | Technical Note | Introduces the DLC design, represented by oracle attestations, that enables contingent settlement on Bitcoin without publishing data on-chain. |
| 2018 | Montoto Monroy, F. J. A. [24] | Bitcoin Gambling Using Distributed Oracles in the Blockchain | Master Thesis | Designs and implements a distributed-oracle betting protocol on Bitcoin, relying on incentive-compatible oracle participation (Orisi-like threshold model). |
| 2019 | Fournier, L. [25] | One-Time Verifiably Encrypted Signatures A.K.A. Adaptor Signatures | Technical Note | Introduces adaptor signatures, a key building block for DLC-based Bitcoin contracts. |
| 2020 | Fournier, L. [26] | How to Make a Prediction Market on Twitter with Bitcoin | Technical Note | Proposes a DLC-based betting/prediction protocol on Bitcoin, coordinated via Twitter, that minimizes rounds and on-chain footprint. |
| 2022 | Le Guilly T. et al. [27] | Bitcoin Oracle Contracts: Discreet Log Contracts in Practice | Conference Paper | Tests DLCs in practice and evaluates practical design choices. |

| 2022 | Shewetark Patel [28] | Transformable Discreet Log Contracts | Master Thesis | Propose TDLCs that allow DLC positions to be transferable before expiration. |
|---|---|---|---|---|
| 2023 | Madathil et al. [29], [30] | Cryptographic Oracle-Based Conditional Payments* | Conference Paper | Introduces Verifiable Witness encryption based on threshold signatures (VweTS) to better support multi-oracle selection and multiple outcomes. |
| 2023 | Hartman et al. [31] | Blockrate Binaries on Bitcoin Mainnet: Formalizing the Powswap Protocol | Whitepaper | Proposes an oracle-less Bitcoin binary derivative using block height and block timestamps as settlement triggers. |
| 2024 | Pierog, J. [32] | Non-Custodial Prediction Markets on Bitcoin | Technical Note | Proposes a non-custodial prediction market design using an LMSR-style market maker implemented via DLCs, highlighting liquidity/exit challenges. |
| 2025 | Han et al. [33] | Vwe2psTS: Supporting Oracle-Based Conditional Payments from Joint Addresses | Conference Paper | Introduces Vwe2psTS, which enables ObC to leverage a single key spend, reducing on-chain footprint and fees while increasing censorship resistance. |

*This paper has two different versions.

## 4. Findings

In the retrieved sample, DLCs appear to be the main post-2015 advancement for Bitcoin L1 oracle mechanisms. After Dryja [23] introduced the DLC concept, subsequent contributions refined the design and applied it to Bitcoin-native use cases. This section first examines Dryja's original proposal and then synthesizes later improvements and applications built on the DLC rationale. The final part discusses other Bitcoin L1 oracle approaches that remain largely conceptual or experimental but are relevant to mapping the broader design space.

### 4.1 Discreet Log Contracts (DLCs).

A key limitation of multisig-based oracle templates on Bitcoin is privacy. On-chain, the script/structure often makes it evident that an escrow-like arrangement exists. DLCs aim to improve privacy and scalability for conditional contracts on Bitcoin, while keeping the oracle dependence as narrow as possible. Dryja [23] argues that DLCs are well-suited to situations in which the outcome of the contract between the parties depends on a value that is publicly known in the future. In practical terms, DLCs make Bitcoin payments contingent on external outcomes while avoiding explicit on-chain oracle messages. The basic idea is that two parties (e.g., Alice and Bob) want to set up a contract, but since they don't trust each other to deliver the funds at a certain point, they rely on the Bitcoin blockchain. Still, they wish for privacy, so they do not want to publish a contract that shows there is an agreement between the two. Finally, they require an oracle to attest to the external outcome that determines settlement, and whatever the outcome, it is executed automatically. DLCs claim to allow parties to execute a contract with these characteristics with the following mechanics.

First, Alice and Bob lock funds in a shared 2-of-2 output, which is basically the funding transaction that locks the money and prepare a set of Contract Execution Transactions (CETs), one for each possible outcome. In practical terms, they pre-sign these CETs in advance and store them off-chain so that once the oracle attests the outcome, the corresponding CET can be broadcast and confirmed without requiring further cooperation between the parties. In addition, they prepare a refund/timeout transaction with a timelock

(time delay), so that if the oracle never attests (or the process stalls), either party can recover funds after a predefined time. Therefore, either the contract settles through the CET corresponding to the oracle-attested outcome, or, if the oracle does not attest or the protocol stalls, the parties can recover funds via a pre-signed refund transaction after a timelock, similar to timeout paths used in Lightning Network constructions.

As anticipated, for the contract to be executed, the two parties also require a third party (e.g., Olivia) to serve as the oracle and verify the outcome of the contract's event object. Unlike multi-sig templates, where oracles may appear directly in the spending conditions, in DLCs, the oracle provides an off-chain attestation, such as a signature, for the realized outcome. The oracle typically announces the event in advance, describing what will be attested, when, and how outcomes are encoded, so that parties can prepare the relevant CETs.

In particular, following Dryja's explanation, Olivia publishes $V = vG$, her long-term public key, and (for each event) a one-time nonce point $R = kG$ committed in advance, together with metadata describing the event (e.g., asset type and closing time). Technically, for each possible outcome $i$, Alice and Bob can precompute the public point $s_i G = R - h(i, R)V$ and use it to construct CET outputs, which include tweaked public keys, e.g., $PubA_i = PubAlice + s_i G$. At maturity, the oracle sets $m$ to the realized value (e.g., 1050) and publishes the corresponding signature scalar $s_m$ produced by signing $m$ under its public key $V$, using the pre-committed event nonce $R$ ($s_m = k - h(m, R)v$. Once $s_m$ is available, the parties can select and broadcast the already-prepared (and pre-signed) CET corresponding to $m$. Revealing $s_m$ also gives the winner the missing secret needed to derive the matching spending key (their own private key plus $s_m$) to spend the payout output.

Because the oracle attestation is off-chain and settlement transactions are designed to look like ordinary spends, DLCs reduce on-chain observability of both the contract structure and the oracle's involvement. The oracle, however, remains a single point of trust as participants must still assume honest reporting. Interestingly, the scheme is designed so that if the oracle equivocates and signs two different outcomes $m_1 \neq m_2$ using the same nonce point $R$, the two signature scalars satisfy a relation of the form

$$s_1 - s_2 = (h(m_2, R) - h(m_1, R))\, v,$$

which allows solving for the oracle's long-term secret key $v$. Hence, any equivocation under the same committed $R$ allows any observer to compute the oracle's private key, creating a strong cryptographic deterrent against double-reporting. Intuitively, however, this mechanism cannot prevent the oracle from signing one false outcome.

In the context of Bitcoin oracles, DLCs therefore represent a shift from on-chain data feeds to off-chain attestation oracles, aiming to improve both scalability and privacy, going beyond the classic multi-signature scheme. Other papers further outline practical implementations of this model in the paragraphs that follow.

**4.2 Extensions to DLCs and VweTS.**

Le Guilly et al. [27] extend Dryja's DLC construction by focusing on implementability and scalability in real deployments, aiming at their standardization. They present an updated DLC protocol based on adaptor signatures that simplifies contract establishment and makes the original punishment design based on time expiration unnecessary. Parties exchange adaptor signatures for all CETs and a refund signature before funding, and at maturity, the oracle attestation allows either party to decrypt exactly one CET signature and broadcast the corresponding settlement transaction.

The use of Schnorr adaptor signatures for DLCs was first proposed by Somsen, building on Poelstra's concept of "scriptless scripts", in a mail discussion between Dryja and Poelstra [34], [35]. Fournier [25] gives a formal foundation for adaptor signatures by modeling them as one-time Verifiably Encrypted Signatures (one-time VES). In simple terms, a party can produce a signature that is verifiably correct but still "locked," and it becomes a normal usable signature only after a specific secret is revealed. Moreover, once the final (unlocked) signature is publicly available, the secret can be recovered from the pair (locked signature, final signature). Fournier revisits the classic VES setting in which a trusted adjudicator decrypts the signature if one party aborts, and adapts the security model to Bitcoin layer-2 protocols, where such a trusted third party is typically not assumed.

As an application to DLCs, Fournier notes that Dryja's original approach can require three on-chain transactions to settle a bet in an uncooperative scenario, whereas using a one-time VES/adaptor technique allows a two-transaction flow by first funding a single joint output, and second by broadcasting the single settlement transaction that becomes valid once the oracle reveals the "unlocking" secret corresponding to the realized outcome.

Finally, Fournier notes that adaptor signatures are straightforward with Schnorr, while an ECDSA version is more fragile and can unintentionally reveal extra information, so pre-Taproot implementations often preferred "semi-scriptless" constructions that rely on simple Bitcoin scripts (e.g., OP_CHECKMULTISIG).

Dryja did not frame DLCs building on adaptor signatures in 2017, probably because Bitcoin at the time supported only ECDSA, while Schnorr (BIP340) became available on Bitcoin only with the Taproot activation in November 2021. Although Malavolta [36] demonstrated the usability of adaptor signatures for ECDSA, as explained by Fournier, they required multiple workarounds for DLCs, making them impractical.

Le Guilly et al. [27] also propose practical extensions that mitigate two major DLC scaling bottlenecks when leveraging multiple outcomes and multiple oracles. For large-range numeric events, they reduce the number of pre-built outcomes by having the oracle attest to the value's digits and by compressing constant-payout intervals into a small set of prefix patterns (e.g., "16___"), substantially cutting contract size at the cost of additional client-side precomputation. They also generalize DLCs to multiple oracles, including threshold settings, and suggest an approach for numeric reports that tolerates small discrepancies by accepting values within pre-defined error bounds (e.g., 2-of-3 or 3-of-5). Importantly, adaptor signatures still provide public auditability since, if an incorrect payout occurs, the attestation can be recovered from the final signature and verified. The authors further report benchmarks and optimizations (e.g., precomputation and parallelization), indicating these extensions remain computationally practical.

Madathil et al. [29], instead, take a different route and argue that some DLC limitations are structural in the multi-oracle and large-outcome constructions. They formalize the concept of oracle-based conditional payments (ObC) and introduce Verifiable Witness Encryption based on Threshold Signatures (VweTS) as the cryptographic core primitive. Intuitively, instead of preparing one adaptor-signature path for every (outcome x oracle-subset) combination, an agent (e.g., Alice) prepares candidate settlement transactions and "locks" the corresponding payment authorization so that it becomes usable only if a sufficient number of independent oracles attest the same outcome. Concretely, the oracles' attestations act as the witness that enables decrypting/recovering the valid payment signature for the realized outcome, after which the agent can broadcast a normal-looking transaction on-chain.  The chain only verifies a standard signature, and no on-chain oracle data needs to be interpreted. VweTS, however, requires spending from 2-of-2 multisignature addresses, which increases on-chain footprint and fees and yields a more distinguishable spending pattern. Han et al. [33] address this deployability concern by introducing Vwe2psTS, which enables ObC from joint addresses using a two-party joint signature that is indistinguishable from a regular single-signature spend. They also report at least a 40% reduction in communication cost relative to VweTS at the cost of higher storage. TVwe2psTS construction also reduces the exposure to discriminatory censorship that targets uncommon transaction forms (e.g., OP_Return war [8])

Compared to DLCs, the VweTS constructions display three main advantages. First, they support threshold multi-oracle settings more naturally because the contract does not need to anticipate which specific subset of oracles will sign. Second, they separate the oracle attestations from the on-chain payment signature mechanism, which makes the approach compatible with multiple signature schemes (including Schnorr/ECDSA/BLS). Third, they are proposed alongside a formal security analysis that should overcome known DLC weaknesses. Unlike threshold DLC constructions, where the off-chain adaptor/encrypted signature material can grow combinatorially because parties must cover every possible t-of-

N oracle subset, VweTS avoids this blow-up and keeps setup essentially linear in the number of oracles.

The trade-off, however, is that VweTS shifts more work off-chain. For small outcome spaces, DLCs can be faster, but Madathil et al. [29] benchmarks indicate VweTS scales better as the number of outcomes and/or oracles grows. Moreover, modern DLC constructions based on adaptor signatures can reduce the on-chain footprint and, depending on the exact implementation, produce settlement transactions that are hard to distinguish from ordinary spends, similarly to Vwe2psTS. Figure 2 illustrates these parallels while comparing the DLC and VweTS workflows.

**Figure 2.** DLCs vs VweTS/Vwe2psTS. Workflow comparison and deployability trade-offs

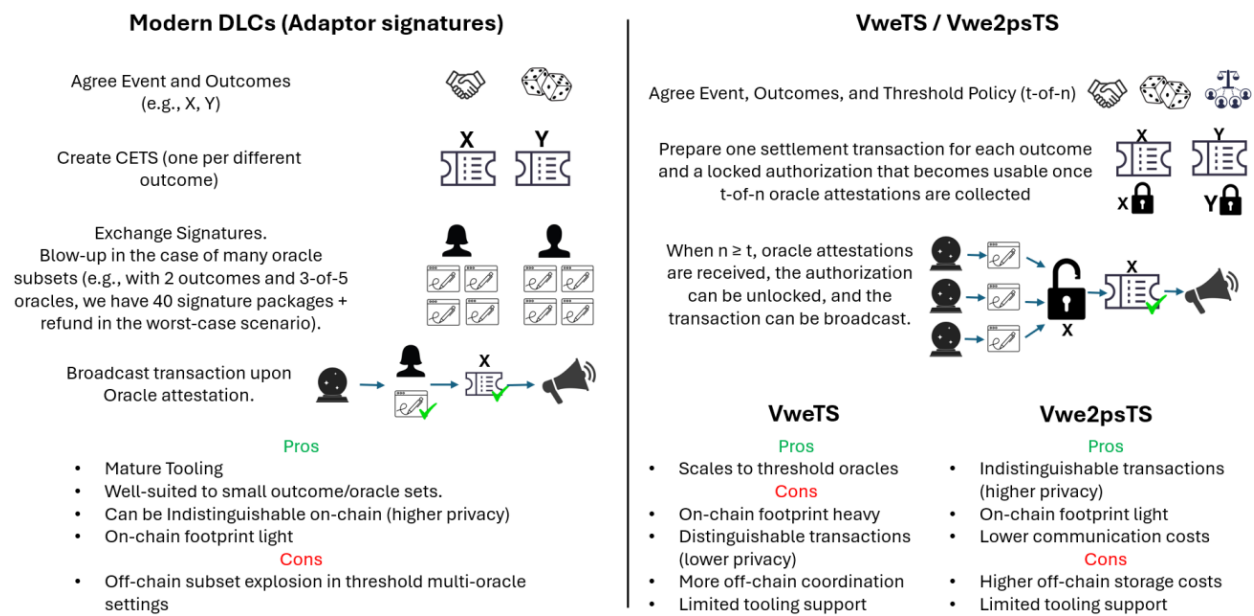

To conclude, it's worth noting that DLCs are not just a research idea; they are backed by an open standard ("dlcspecs") and by practical toolkits that cover the whole workflow, from negotiating the contract and building/signing CETs to connecting oracles and executing the payout. In particular, the DLC Dev Kit integrates these components into a modular framework, reducing the engineering burden of building DLC-enabled applications and enabling other proposals and integrations to be based on DLCs beyond the field of oracles [37], [38], [39]. By contrast, VweTS-based constructions primarily remain a research/prototype direction offering a way to scale to threshold multi-oracle settings, but they have not yet converged into a comparably standardized and widely adopted wallet/tooling stack.

## 4.3 Applications of DLCs for betting and prediction markets.

Fournier [26] proposes a lightweight DLC-based "prediction market on Twitter" that uses a public message board to discover counterparties and exchange setup messages for binary, winner-takes-all events, aiming at only two rounds of communication and two on-chain transactions (funding and settlement). Alice posts a public proposal specifying the event (oracle + event identifier), her stake, the UTXOs she intends to spend, a change address, and a fresh public key. Potential counterparties reply with offers in which only their public key is visible, while the remaining parts of the offer, including the chosen side and input witnesses, are encrypted under a Diffie–Hellman shared secret derived from the two parties' keys. From an accepted proposal/offer pair, the parties deterministically construct a fund transaction whose main output is a 1-of-2 spending condition using two tweaked public keys that incorporate oracle outcome commitments. In practical terms, one key becomes spendable only if the oracle later reveals the discrete log corresponding to the winning outcome, thereby allowing the winner to claim the pot once the oracle attests off-chain. The paper also highlights several protocol-level fine distinctions. It notes that key-tweaking constructions originally proposed in Dryja [23] can invite rogue-key-style attacks, allowing a malicious party to choose a public key to cancel an oracle outcome key and thus derive a spending key without waiting for the oracle. Therefore, Fournier sketches why the protocol's Diffie–Hellman structure makes such attacks implausible under the Knowledge of Exponent Assumption. Furthermore, in terms of privacy, coordinating the bet via a public Twitter board creates an inherent leakage because Alice's posted inputs and change address can be linked to the on-chain fund transaction. Acknowledging this limitation, Fournier's design targets "winner privacy" and "offer privacy" by encrypting offer details and randomizing/permuting the on-chain keys so observers cannot trivially infer which offer was accepted or which party won. Finally, Fournier discusses practical market-manipulation risk, labeled post-event acceptance, for which efficient countermeasures have yet to be demonstrated. Alice could, in fact, wait until the outcome is known and then accept only favorable offers. In this scenario, the only plausible defense for counterparties is to cancel offers by spending one of their inputs. This is quite cumbersome to solve because Bitcoin supports timelocks that make transactions "invalid-until" but not "valid-until" a certain time. Therefore, the paper suggests a workaround in which Alice pre-commits to a timelocked spend of one proposal input that becomes valid before the event outcome time, enabling proposals not intended to be fulfilled to be cancelled with a single broadcast.

Another useful contribution on DLC implementation is Pierog's work [32], which addresses the liquidity constraints that limit DLC-based prediction markets. To tackle this problem, he proposes a non-custodial prediction market on Bitcoin that combines an LMSR-style automated market maker with DLCs, treating each trade as a binary DLC between a trader and the market maker rather than issuing tokenized "shares." In this design, buying $Q$ shares requires the trader to lock the share cost $K$ in sats as DLC collateral while the market maker commits the remaining collateral $R = 100Q - K$, so that the total value locked *(K+R =*

*100Q)* equals the eventual payout capacity, considering that when the event is known, each share receives either 0 or 100 sats. Considering a binary (y/n) event, AMM maintains a market state $q = (y, n)$ and quotes prices using a modified LMSR cost function with a commission parameter $\alpha$. Given an average purchase price $p = K/Q$, the resulting DLC (liquidity) ratio satisfies $K/R = K/(100Q - K) = p/(100 - p)$, matching the share price ratio $p: 100 - p$ at the time of purchase. Intuitively, pricing and inventory accounting are managed off-chain. Settlement is enabled instead on Bitcoin L1 once the oracle publishes (off-chain) its attestation signature for the realized outcome, which unlocks the corresponding CET.

Importantly, Pierog highlights another limitation of DLC-based prediction markets. Once a trader enters a position, the BTC used as collateral is locked in the DLC until the event is resolved, similar to locked-margin contracts, so the trader cannot "sell" the position early. A prediction market is instead useful "before" the event ends, because people trade in and out as their beliefs change. DLCs, by default, don't let parties do that. The article suggests that a functional DLC-based market should include an exit/swap mechanism that allows participants to transfer or close positions before the event resolves, since otherwise collateral remains locked and trading, along with price discovery, may largely stop.

A complementary line of work tackles this limitation directly by making DLC positions transferable. In his thesis, Patel [28] proposes Transformable DLCs (TDLCs), where the original counterparties pre-arrange additional "transformation" transactions that allow a third party to swap into the contract mid-way. Intuitively, the exiting trader and the entering trader agree on a price, execute a transformation step that replaces the exiting party's key in the contract, and the position continues with unchanged oracle settlement logic, without requiring the remaining counterparty to be online at the moment of the transfer. The thesis further introduces a stronger variant, Truly Transformable DLCs (TTDLCs), that allow a position to be traded multiple times among a predefined set of participants, thereby approximating secondary-market trading of cash-settled futures on Bitcoin. While these designs improve "exit" and liquidity, they also introduce added setup complexity, especially for TTDLCs, and, like standard DLCs, still require full collateral to be locked until final settlement.

### 4.3.1 An oracle-less proposal for Bitcoin-based derivatives

Concerning Bitcoin-based betting, it is also worth noting a complementary line of work that avoids external oracles by relying on chain-observable variables. Hartman et al. [31] formalize PowSwap as Blockrate Binaries, a Bitcoin-native binary contract in which the payout depends on whether a chosen block-height or time threshold is reached first, using pre-signed settlement transactions and timelocks. This design de facto removes oracle trust assumptions, but it introduces several practical challenges. If the two thresholds are reached very close in time, the outcome may hinge on a confirmation race between competing settlement transactions, amplifying fee-bumping dynamics and making the protocol sensitive to transaction-delay attacks (e.g., pinning or censorship). Moreover, the

"time" condition relies on block timestamps, which are chain-native but not perfectly objective and can be influenced by miners within protocol limits. The authors therefore position Blockrate Binaries as an oracle-less alternative suited to a narrow class of events, and they note that broader usability depends on robust peer-to-peer counterparty discovery and market infrastructure.

### 4.4 Bitcoin as public oracle registry (Orisi extension)

Given the limited research in this field, it is reasonable to also present a master's thesis that develops a protocol-style proposal for a non-custodial prediction market on Bitcoin and proposes its oracle design. The aim of the protocol is to reduce trust in counterparts such as betting sites. The design involves two players who may lock the wager via Bitcoin transactions, and a set of oracles that subsequently determine how the funds are allocated. In practical terms, the thesis proposes using the Bitcoin blockchain itself as a public oracle registry, where anyone can register as an oracle by publishing a recognizable transaction that includes registration data, via an OP_RETURN payload, enabling players to compile a list of available oracles from-chain. The two players then agree on parameters such as the wager amount, the number of oracles to use, oracle fees, penalties, and timeouts, and select a subset of oracles via a distributed coin-tossing procedure, ensuring that the selected set is random rather than handpicked and reducing the risk of collusion between players and oracles. At the bet setup, the players broadcast a "Bet Promise" transaction that publicly states the bet description and the IDs of the invited oracles. Invited oracles accept the job by submitting a deposit, which may be forfeited if they fail to participate properly. Once enough oracles have enrolled, the players publish the Bet transaction that funds the wager and sets aside the oracle reward. The transaction is structured to unlock the funds upon the winner's signature, with sufficient secrets revealed by the oracles in their votes on the chain. Alternatively, with the loser's signature plus the necessary elapsed time. After the real-world event happens, each oracle submits its vote on the chain to collect the reward. When a threshold (e.g., *m-of-n*) of oracle votes converges, the winning player can redeem the locked funds, and after an additional timeout, the players can confiscate deposits from oracles who did not vote, incentivizing both honesty and liveness.

The author of the thesis explicitly acknowledges similarities between its design and the Orisi protocol, while highlighting crucial differences. Regarding oracle selection, the Orisi protocol proposes a trusted oracle list that signs transactions alongside the players. The thesis instead proposes selecting public oracles via blockchain and removing their signatures from the multisig script escrow, enabling smaller transactions. The practical feasibility of this design, however, raises some concerns. Although theoretically feasible, as it leverages Bitcoin's building blocks (timelocks, P2SH-style scripts, OP_RETURN metadata), the proposal heavily relies on on-chain transactions. As recognized by the author, the cost of this protocol for a simple bet in 2019$ was nearly 60$ de facto making bets of lower value infeasible. With Bitcoin's price rising considerably, the cost of the process would rise as well, making real implementations highly impractical. Finally, the privacy of the oracles is effectively forfeited, as all their votes are on-chain, and the lack of a barrier to new oracle entries greatly facilitates sybil schemes.

**4.5 Craig Wright's Bitcoin oracle.**

Another project worth noting comes from Craig Wright, a controversial figure in the Web3 community due to his claims to be Satoshi Nakamoto. In 2016, Wright proposed nChain [22], which he claimed was an oracle that enabled Bitcoin scripts to become Turing-complete. The oracle is framed as an IFTTT-like automation engine whose "If" conditions may depend on blockchain events or external data (time, weather, IoT, news), and whose "Then" actions may include Bitcoin payments or off-chain operations, with transaction metadata recording an audit trail of each iteration.

As a concrete oracle application, the paper envisions an automated vote-counting bot in which the oracle first distributes voting tokens represented by Bitcoin key pairs funded with a small amount of BTC (one vote). Then it publishes the list of eligible public keys, destroys the identity/token mapping, and runs a repeat loop that processes the list of addresses, counting unique "Yes" votes until the list is exhausted. If the count reaches a threshold (e.g., 57 "Yes" votes) before a deadline, the bot triggers a payment, and it discloses the code hash along with a link, or other reference, to where the exact code can be retrieved, so third parties can recount the votes and verify the audit trail.

From a feasibility standpoint, the proposal appears robust, as an oracle of this kind could, in practice, enable a loop on the Bitcoin blockchain to execute operations until a specific condition is met. The evident issue is that this design would inherently shift trust from Bitcoin to the oracle module, making it a single point of failure not only for external data gathering but also for off-chain transaction execution.

The next section discusses these findings in light of prior work. Figure 3 summarizes the evolution of Bitcoin oracle designs and relevant protocol milestones.

**Figure 3.** Timeline of modern Bitcoin oracle development and integrations

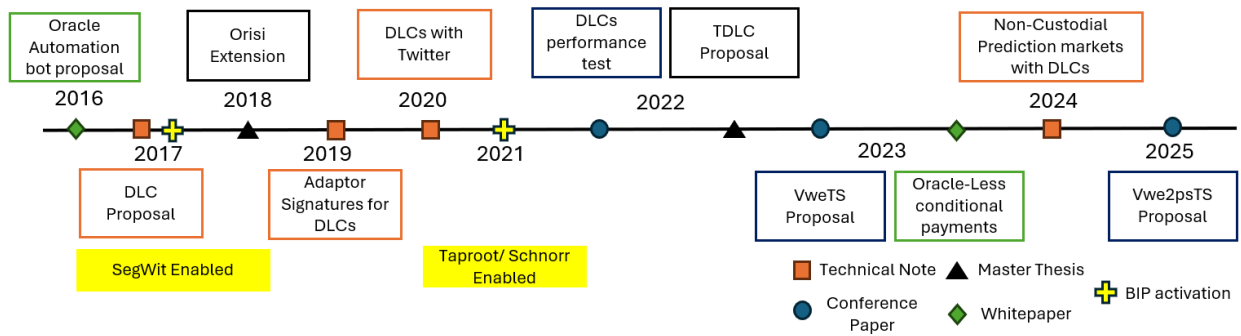

## 5. Discussion

This review updates the post-2015 landscape of Bitcoin oracles. Despite limited coverage in indexed journals, innovation has continued, often documented in developer-oriented manuscripts rather than academic outlets, and shaped by Bitcoin's architectural and cost constraints. The following discussion extracts three recurring patterns from the retained contributions.

### 5.1 Architectural shift from on-chain disclosure to off-chain attestation

The most evident post-2015 pattern is a shift in how oracle information is represented and consumed. Early Bitcoin oracle designs often conceptualized the oracle as an entity that publishes a statement that becomes visible on-chain, or as an explicit arbiter that directly co-signs a payout transaction. Under this template, the oracle identity and action are relatively transparent. The oracle signature, or on-chain message, serves as the trigger that selects the finalized contract branch. Although conceptually straightforward, this model creates recurring debate in the Bitcoin community. Complex On-chain transactions are not only expensive for users but also data-intensive, at the expense of the entire community, raising questions about their impact on long-term sustainability. On-chain statements also create privacy leaks for the contracting parties and the oracle operator, which the chain's rationale seeks to minimize.

Modern Bitcoin oracle constructions shift from a publication-centric approach toward an attestation-centric approach. The oracle's contribution is an off-chain attestation that enables settlement without requiring the oracle to appear as a distinct on-chain actor. The DLC design provides the clearest example of this approach. In this model, the oracle does not broadcast a transaction and, in principle, should not even know whether its attestation triggers one. The oracle's role becomes closer to that of a generic fact signer, whose signatures can be reused as inputs for many contracts, thereby reducing the on-chain impact of oracle-based contracts to the settlement pattern, which remains composed solely of funding and closing transactions with a fallback refund path.

This shift is not just a technical change; it also reshapes how oracles are conceived and organized in Bitcoin-native contracts. Reducing the oracle's on-chain visibility makes real-time monitoring less direct, and shifts verification and accountability to off-chain elements such as published attestations, logs, or other service-level disclosures. In the Dryja [23] approach, parties are assumed to select an oracle they already trust for the relevant event. If it finally turns out to be untrustworthy, then they don't have many countermeasures other than simply selecting another one. To tackle this challenge, Le Guilly et al. [27] introduce thresholding to mitigate the effects of oracle non-availability or misreporting. However, because the mechanism is largely off-chain, transparency and auditability are not automatically provided by the blockchain and depend on how oracle attestations are disclosed and verified. Overall, developer effort appears more focused on solving implementation challenges than on theorizing the oracle's role in terms of liveness, key management, long-term reliability, and standardization.

Considering the oracle literature as a whole, the identified research stream reflects an intense cryptographic "battle of minds" which is technically sophisticated and intellectually impressive, but with a long-term practical impact difficult to evaluate ex ante. Prior research on Bitcoin oracles suggests that part of the historical migration toward easier-to-program chains was also driven by the cryptographic and engineering complexity surrounding Bitcoin-based designs. In this respect, proposals that further increase complexity may struggle to achieve broad practitioner adoption and, consequently, to generate sustained practical relevance. Building on earlier work in this area, secure oracle-based transactions on Bitcoin and other chains may benefit from more interdisciplinary research rather than purely technical refinement, particularly by incorporating philosophical, governance, and game-theoretic perspectives that remain underdeveloped in many modern Bitcoin oracle constructions.

**5.2 Knowledge production in Bitcoin oracles.**

As noted above, the pre-2015 review necessarily relied on non-academic sources, since peer-reviewed work on Bitcoin oracles was largely absent at the time. Despite blockchain becoming an established research field, the present study encountered similar limitations when focusing specifically on Bitcoin-native oracles. Therefore, a second pattern that emerges from this study concerns how Bitcoin oracle knowledge is produced and disseminated. Across both the earlier review and this update, the most influential contributions often appear as developer manuscripts, preprints, or technical proposals, sometimes authored by academics, sometimes by practitioners, but frequently not framed as conventional journal articles. This pattern has important implications for how blockchain research is reviewed and synthesized.

This divergence suggests that Bitcoin-oracle innovation and research are primarily devoted to implementation and ecosystem adoption rather than to securing scientific prestige. Although commendable, this poses a clear methodological challenge for a research field that aims to document and interpret technical evolution. Standard systematic literature reviews, which privilege peer-reviewed and indexed venues, may underrepresent the most impactful developments in implementation-driven domains. Arguably, if this pattern can be extended to other Bitcoin or blockchain-related implementations, this supports the use of multivocal review approaches, especially where key contributions are frequently published outside journal venues. Notably, reviews in other blockchain domains already adopted MLR as a methodology, but the rationale of adoption was recognized in the early stages of research and not in a systematic literature production bias [40], [41], [42], [43]. Accordingly, reviews of Bitcoin-native oracle mechanisms should then incorporate grey literature using established multivocal review procedures.

**5.3 Limited Bitcoin oracle application landscape.**

A third evident pattern emerging from this research concerns the application focus. Beyond early conceptual examples such as Bitcoin inheritance proposed by Hearn [2], most subsequent oracle proposals were developed in the prediction market/betting context, for which Reality Keys, Orisi, and Truthcoin were famous examples [5], [6], [44]. The collected sample in this study also disproportionately clusters around betting, gambling, and prediction markets. This clustering, however, is not accidental, as discrete-event contracts align naturally with the type of oracles developed on Bitcoin. In Bitcoin-native settings, the most straightforward oracle question is often binary, such as: "Did event X occur?", or at least discretized into a small number of outcomes. In fact, despite Dryja's initial proposal arguably overestimating the feasibility of supporting a broad outcome space, including multiple price points, subsequent formulations in our corpus tend to favor binary outcome structures, even though Le Guilly et al. [27] and Madathil et al. [29] also proposed ways to simplify a larger outcome space.

As Pierog [32] also notes, the main challenge for Bitcoin-native prediction markets is not only oracle design but the market microstructure required for continuous trading and exit liquidity. In prediction markets, traders buy and sell contingent claims based on their beliefs, and market prices can aggregate dispersed information into forecasts of uncertain events [45]. Many practical implementations rely on automated market makers, most notably Hanson's logarithmic market scoring rule, to provide continuous liquidity even when order flow is thin [46]. In contrast, on Bitcoin L1, the lack of a native, low-cost, composable exchange layer makes secondary trading and dynamic price adjustment more difficult, so designs either shift the market engine to separate execution environments, such as Truthcoin's proof-of-work sidechain with an LMSR market maker or implement trading through contract-based constructions such as DLC-based positions mediated by a market maker, at the cost of increased collateral lock-up and reduced liquidity. However, the approach based on DLCs, although it ideally serves as a workaround for Bitcoin's L1 limitations, can make trading operations cumbersome, potentially undermining price discovery and the extent to which prices reflect participants' beliefs. Although the Patel [28] extension elaborates on this concept and attempts to address this issue, it inevitably increases the complexity of DLCs, thereby impacting market price discovery.

In light of reducing complexity, the thesis presented by Montoto Monroy F.J.A. [24], for example, although potentially costly to implement, framed Bitcoin simply as a betting platform, considering binary outcomes and limited interactions. Also, Fournier's [26] proposal to leverage Twitter for DLC-based contracts appears more naturally aligned with the betting context than with full-fledged prediction markets, given his explicit claim of allowing limited player interactions. The point is not that Bitcoin cannot do more in theory, but that on L1 it becomes expensive and hard to scale beyond discrete events. A wider set of oracle-driven applications is therefore more realistic when market logic and data processing move off-chain. Considering Bitcoin's longevity and the fact that, in our sample, Bitcoin-native oracle applications concentrate on discrete, often binary, event resolution, it is reasonable to hypothesize that Bitcoin L1 is best suited to attestation-based settlement of discrete outcomes, while richer oracle-driven functionality involving continuous feeds, composable markets or complex state machines is more realistically achieved through off-chain integrations or adjacent execution environments. Table 2 summarizes the identified modern Bitcoin oracle patterns.

**Table 2**. Legacy vs DLC-based oracle patterns on Bitcoin.

| Characteristic | Legacy Bitcoin Oracles | DLC-based Bitcoin Oracles |
| --- | --- | --- |
| On-chain Interaction | Oracles are the actual triggers that execute the correct contract branch. Contract data is directly stored on-chain | Oracle attestations are produced off-chain, so on-chain settlements look like ordinary spends. Neither the oracle message nor the retrieved data are published on-chain. |
| Privacy | Oracle involvement and contract structure are typically observable on-chain. Parties may remain pseudonymous, but linkage between counterparties/oracle and contract execution is generally easier. | The oracle attestation is off-chain, and contracts can be settled without revealing oracle activity on-chain. However, privacy still depends on off-chain communication and operational metadata. |
| Trust | Oracle behavior is more directly observable and can be monitored through on-chain evidence. Multi-oracle and threshold schemes can reduce reliance on a single source, though they may increase complexity/costs. | Oracles are selected by the counterparties, and accountability relies more on off-chain transparency, including published attestations/keys, uptime, and reputation. The rationale is that users choose the oracle(s) they trust. Multi-oracle thresholding can mitigate downtime/misreporting, but requires a type of coordination that is still under debate. |

**5.4. Bitcoin vs. EVM-like chains oracle designs.**

The above-discussed patterns also help explain why the Bitcoin oracle landscape differs from that of EVM chains. On EVM platforms, smart contracts can store data and run complex rules directly on-chain. This enables a common oracle architecture in which an oracle network continuously updates an on-chain feed. The feed then becomes shared infrastructure, and other contracts compose with it by reading a standardized interface [47], [48]. Consequently, research on EVM-oracles often emphasizes robust aggregation, incentive design (staking/slashing), latency, resistance to manipulation, and governance of shared feeds [49], [50]. Oracles are therefore treated as persistent infrastructure for highly composable applications.

Bitcoin's base layer supports a different equilibrium. Because Bitcoin Script is intentionally constrained and on-chain interaction is costly, Bitcoin oracle designs tend to minimize on-chain "oracle state." Instead of building shared, continuously updated feeds on Bitcoin L1, many proposals rely on off-chain attestations that unlock settlement paths in isolated contracts. This makes Bitcoin oracle usage more "per-contract" and more "event-driven" than "continuous." Moreover, the latest constructions represented by DLCs explicitly aim to make settlement transactions indistinguishable from ordinary spending, which prioritizes privacy and fungibility, values that are less central in many EVM oracle designs, where transparency and shared state are often accepted trade-offs for composability.

This comparison provides us with two further implications for future research. First, it suggests that importing EVM-style oracle assumptions into Bitcoin can be misleading, as the relevant design space is not "how to maintain a feed on-chain," but "how to bind attestations to settlement with minimal on-chain footprint and limited information leakage." Second, it indicates that Bitcoin oracle progress should be evaluated using criteria aligned with Bitcoin's constraints, such as transaction count, round complexity, liveness under partial participation, privacy, and the feasibility of multi-oracle redundancy without requiring a conditional on-chain aggregation contract. Table 3 summarizes the findings of this study, organized by research questions.

**Table 3**. Review Findings

| RQs | Findings | Discussion Takeaway |
|---|---|---|
| RQ1 | After 2015, Bitcoin oracle innovation is largely dominated by the DLC approach, marking a shift from on-chain disclosure toward off-chain attestations that enable settlement. DLCs are also the only approach in our sample backed by tooling and practical demonstrations, while extensions of older protocols and off-chain automation proposals remain largely conceptual. | Bitcoin L1 oracle designs tend to prioritize discrete-event settlement, privacy, and a minimal on-chain footprint. Continuous-feed and highly composable oracle functionality, common on EVM chains, is harder to replicate on Bitcoin L1 and is more naturally achieved via off-chain or Bitcoin-adjacent layers. Research on modern Bitcoin oracles also prioritizes technical improvements rather than philosophical and game-theoretical approaches. |
| RQ2 | DLC-related development is primarily developer-driven (often by Lightning Network developers), with contributions from both practitioners and academics. | Because many influential Bitcoin-oracle contributions appear outside journals, reviews in this domain should use multivocal approaches (e.g., grey-literature–inclusive protocols) to reduce outlet-related biases. |
| RQ3 | While proposals present broad potential applications, the most concrete and repeatedly documented implementations in our sample concentrate on prediction-market/betting settings with discrete or binary outcomes, highlighting the importance of liquidity and exit mechanisms. Interestingly, within recent Bitcoin integrations, an oracle-less alternative is also proposed. | This suggests that Bitcoin L1 can realistically support only relatively simple, discrete oracle applications. More advanced features, such as data feeds, aggregations, or complex contract logic, work better off-chain or on bitcoin-adjacent layers. Alternatively, if the application-specific data is already available on-chain, oracle-less contracts may also be explored. |

**Conclusion**

Overall, the post-2015 landscape suggests a field that is narrow in indexed academic coverage but richer in protocol evolution. The strongest convergence is toward attestation-based oracle interfaces such as DLCs, complemented by parallel lines exploring multi-oracle voting and off-chain execution optimizations. This convergence clarifies why Bitcoin oracle research often appears in the grey literature, as innovation is deeply dependent on deployability and engineering constraints. At the same time, application designs reveal that the remaining bottlenecks are often economic and user-experience driven, focusing on liquidity, exit, and coordination strategies rather than purely cryptographic. Overall, the reviewed implementation patterns suggest that Bitcoin-oracle-based contracts are currently better suited to bespoke, high-value agreements between sophisticated counterparties than to mass-market, composable DeFi applications. This is mainly due to the complexity of the setup, the need for coordination around oracle attestations, and the requirement to lock collateral until settlement.

The present study has limitations, primarily due to the scarcity of indexed academic studies on Bitcoin-native oracles and the difficulty of retrieving relevant materials through keyword searches, where "oracle" has multiple meanings. Future work should therefore complement bibliographic searches with qualitative evidence, including interviews with Bitcoin oracles and application developers, to better understand the challenges overcome, the constraints that persist, and how protocol limits and broader development norms in the Bitcoin ecosystem shape design choices. Finally, the oracle-less design proposed by Hartman et al. [31], despite practical challenges, highlights that Bitcoin-based conditional transactions can sometimes be achieved without external attestations when the relevant condition is fully chain-observable. This suggests a complementary research direction focused on "oracle-less" contracts, identifying which event classes can be expressed safely using only on-chain data and the trade-offs they entail.